\RequirePackage{fix-cm}
\documentclass[smallextended]{svjour3}       
\smartqed  
\usepackage{graphicx}
\usepackage{amsmath}
\usepackage{amssymb}
\usepackage{natbib}
\usepackage{aas_macros}
\begin{document}

\title{Hydrogen dominated atmospheres on terrestrial mass planets: evidence, origin and evolution
}

\titlerunning{H$_2$ atmospheres on terrestrial mass planets}        

\author{J. E. Owen \and I.F. Shaikhislamov \and H. Lammer \and L. Fossati \and M. L. Khodachenko
}

\authorrunning{Owen, J. E. et al.} 

\institute{J. E. Owen \at
              Astrophysics Group, Blackett Laboratory \\
              Imperial College London \\
              Prince Consort Road \\
              London, SW7 2AZ, U.K.\\
              \email{james.owen@imperial.ac.uk}           
           \and
           I.F. Shaikhislamov \at
               Institute of Laser Physics SB RAS\\
               Novosibirsk\\
               Russia
            \and 
            H. Lammer, L. Fossati \& M. L. Khodachenko \at
            Space Research Institute\\
            Austrian Academy of Sciences\\
            Schmiedlstrasse 6\\
            A-8042 Graz\\
            Austria
}

\date{Received: date / Accepted: date}

\maketitle

\begin{abstract}
The discovery of thousands of highly irradiated, low-mass, exoplanets has led to the idea that atmospheric escape is an important process that can drive their evolution. Of particular interest is the inference from recent exoplanet detections that there is a large population of low mass planets possessing significant, hydrogen dominated atmospheres, even at masses as low as $\sim 2$~M$_\oplus$. The size of these hydrogen dominated atmospheres indicates the the envelopes must have been accreted from the natal protoplanetary disc. This inference is in contradiction with the Solar System terrestrial planets, that did not reach their final masses before disc dispersal, and only accreted thin hydrogen dominated atmospheres. In this review, we discuss the evidence for hydrogen dominated atmospheres on terrestrial mass ($\lesssim$ 2~M$_\oplus$) planets. We then discuss the possible origins and evolution of these atmospheres with a focus on the role played by hydrodynamic atmospheric escape driven by the stellar high-energy emission (X-ray and EUV; XUV). 
\end{abstract}

\section{Introduction}
The picture in which planets grow from small, $\sim 1~\mu$m sized, dust-particles in the star's natal protoplanetary accretion disc is the paradigm in which the majority of planets form \citep[e.g.][]{Lissauer1993}. This protoplanetary disc is initially composed of roughly 1\% solids and 99\% hydrogen/helium (H/He) gas by mass \citep[e.g.][]{Armitage2011}. In the core-accretion paradigm, terrestrial and gas-giant planets begin forming in a similar fashion. Mutual collisions between solids lead to growth and eventually, via a debated mechanism, the formation of 1-100 km-sized planetesimals \citep[e.g.][]{Chiang2010,Birnstiel2016}. At $\sim$ AU separations, within about 10$^5$ years, mutual collisions and runaway accretion of planetesimals produce larger planetary embryos with sizes of the Moon- to Mars, some of which grow further via collisions to terrestrial planet sizes \citep[e.g.][]{Morbidelli2012}. 

At large separations in the Solar System, this collisional growth proceeded to the point where the planetary embryos became massive enough to accrete large amounts of H/He from the nebula, and eventually became gas giants \citep[e.g.][]{Pollack1996,Helled2014}. In the case of the Solar-System terrestrial planets, it is thought that the solid surface density was such that the embryos were not able to reach large enough masses to accrete a significant amount of gas before the nebula dispersed \citep[e.g.][]{Morbidelli2012}. Thus, there was growth after the gas disc dispersed, perhaps through a giant impact phase. \citet{Venturini2020} provides a recent summary of planet formation in a companion article to this review. 

Exoplanet discoveries have taught us that the distinction between planets possessing large hydrogen/helium atmospheres and those which do not is blurred. The discovery of super-Earths and mini-Neptunes \citep[e.g.][]{Rivera2005,Borucki2010} have indicated that planets close to their star (with semi-major axes $<1$~AU) show diversity in terms of the presence or absence of a H/He atmosphere.  Even planets as low-mass as $\sim$ 2~M$_\oplus$ \citep{Lissauer2011,Masuda2014,Xie2014} have been inferred to possess a voluminous H/He atmosphere. In this review, the meaning of ``voluminous'' specifically refers to the case where the amount of H/He in a planet's atmosphere is large enough to be detectable through the measurement of a planet's density. Therefore distinguishing planets that have $\ll 0.1$\% of the planet's mass in a H/He atmosphere against those that have $\gtrsim 1$\%. In fact, many known exoplanets have a H/He atmosphere which contains $\sim 1\%$ of the planet's mass \citep[e.g.][]{Lopez2014,Wolfgang2015,Jankovic2019}. 

Due to observational biases in detecting exoplanets, many of the observed exoplanets are close to their host star, and as such, intensely irradiated. Further, the majority of known exoplanet host stars are several billion years old \citep[e.g.][]{bonfanti2015,McDonald2019,Berger2020}. During the star's early life, its X-ray and UV output was significantly higher than its value after several billion years \citep[e.g.][]{Ribas2005,Jackson2012,Tu2015}. These high-energy photons heat the planet's H/He dominated atmospheres to sufficiently high temperatures (of the order of 10$^4$\,K) to drive hydrodynamic atmospheric escape \citep[e.g.][]{Lammer2003,Yelle2004,MurrayClay2009,Owen2012,Koskinen2014}. Over the course of the planet's early lives, atmospheric escape can remove partly or entirely the planet's atmosphere \citep[e.g.][]{Lopez2013,Owen2013,Jin2014,Lammer2014,Erkaev2016,Kubyshkina2018_grid}. In fact the observed radius-gap \citep{Fulton2017,Fulton2018,VanEylen2018} that separates terrestrial-like ``super-Earths'' which are believed not to possess voluminous H/He atmospheres from those ``mini-Neptunes'' which are believed to have voluminous H/He atmospheres is likely caused by atmospheric escape \citep{Owen2017,Lehmer2017,Jin2018,Wu2019,Owen2019}. 
When this escape driven evolution is taken into account it is currently possible to explain the vast majority of close-in super-Earth planets through an evolutionary pathway in which they accreted a voluminous hydrogen/helium atmosphere which they have then lost \citep[e.g.][]{Owen2017,Kubyshkina2019a,Rogers2020}. However, the quantitative details are still uncertain, with most planet formation models over-predicting the amount of H/He a planet can accrete \citep{Jankovic2019,Ogihara2020}. 

Since current exoplanet observations are typically limited to planets with masses and radii above that of the Earth and that are also close to their parent stars, it is unclear whether this trend can be extrapolated towards lower, Solar-System like terrestrial mass planets. In this review we explore this question: Namely, can planets with masses similar to the Solar-System terrestrials (e.g. $M_p\sim0.1-2$~M$_\oplus$) accrete and retain a voluminous H/He atmosphere, and under what conditions? 

Whether or not terrestrial mass exoplanets accrete voluminous H/He atmospheres from the nebula, which they then retain or lose, obviously has important consequences for habitability \citep[e.g.][]{Lammer2014,Luger2015,OM16} and for the composition of any secondary atmosphere. In this review, we discuss the theoretical evidence from models of planet formation and evolution, as well as what the Solar-System terrestrial planets tell us. We use this evidence to speculate about the origin and evolution of hydrogen dominated atmospheres on terrestrial planets. 

\section{Accretion of H/He from the nebula}
A fundamental question regarding the accretion of H/He from the nebula is how massive a protoplanet needs to be to gravitationally bind gas. The answer to this question is relatively simple: a planet needs to be massive enough such that any gas particle that reside on the planet's surface does not have sufficient thermal energy to escape the potential well of the planet. The radius at which gas particles have sufficient thermal energy to escape a point mass potential is roughly given by the ``Bondi-radius'' \citep[e.g.][]{Ikoma2006}:
\begin{equation}
    R_B=\frac{\gamma-1}{\gamma}\frac{GM_p}{c_s^2}\,,
\end{equation}
with $M_p$ the planet's mass, $\gamma$ the ratio of specific heats, and $c_s$ the sound-speed of the gas. Therefore, for a planet to gravitational bind an atmosphere we require the planet's radius ($R_p$) to be smaller than its Bondi-radius, or quantitatively \citep[e.g.][]{Ginzburg2016, Massol2016}:
\begin{equation}
M_p\gtrsim10^{-2}\,{\rm M}_\oplus\,\left(\frac{T}{300\,{\rm K}}\right)^{3/2}\left(\frac{\rho}{5~{\rm g~cm}^{-3}}\right)^{-1/2}\,,
\end{equation}
for $\gamma=7/5$. Therefore, planets reaching a mass comparable to any of the terrestrial planets in the Solar-System before the gas disc disperses have no trouble actually accreting a H/He dominated atmosphere. 

Therefore, the question of whether terrestrial mass planets generally possess a hydrogen dominated atmosphere is not a question of whether they can actually accrete one. But rather: how massive a hydrogen dominated atmosphere do protoplanets accrete and then retain for billions of years? 

Below we discuss the expectations from theory and the evidence from early Venus and Earth. 

\subsection{Theoretical expectations}\label{sec:acc}

The question of how much gas a terrestrial mass core can accrete from the nebula is the problem of core accretion. In classical models of core-accretion \citep[e.g.][]{Pollack1996,Rafikov2006,Piso2015}, primarily aimed at forming giant planets, the proto-planets have sufficiently high solid accretion rates that heat-up the protoplanets' atmospheres such that terrestrial planet cores accrete negligible atmospheres \citep[e.g.][]{Rafikov2006,Lammer2014}. 

However, with the detection of low-mass ($\gtrsim 2$~M$_\oplus$) planets\footnote{Terrestrial mass exoplanets $\lesssim 1$~M$_\oplus$ that are sufficiently far from their stars to retain a H/He atmosphere (See Section~\ref{sec:loss}) are too far to have their masses measured with the RV method with current instrumentation.} that possess voluminous H/He atmospheres \citep[e.g.][]{JontofHutter2016}, the question of atmospheric accretion in the core-accretion paradigm for lower-mass planets has been re-investigated. \citet{Ikoma2006} and \citet{Hori2010} showed that for sufficiently low solid accretion rates, such as those that might occur once a planet depletes its feeding zone, Earth mass and lower cores could accrete voluminous H/He atmospheres.  Further, \citet{Lee2014,Lee2015,Lee2016,Ginzburg2016} argued that if the heating from core-accretion was sufficiently high to prevent accretion of an atmosphere for terrestrial mass protoplanets, then the core-accretion rate would be sufficiently high that it would not remain a terrestrial mass core over the several million-years nebula lifetime. Therefore, if protoplanets are to grow to terrestrial masses ($0.1-2$~M$_\oplus$), but not further before the disc disperses, the solid accretion rates onto the core must be low enough that the core mass doubling time is no shorter than a few Myr. Therefore, these low solid accretion rates are typically unimportant for the thermodynamics of atmospheric accretion \citep{Ikoma2006, Lee2014}. 

\begin{figure}
    \centering
    \includegraphics[width=0.7\textwidth]{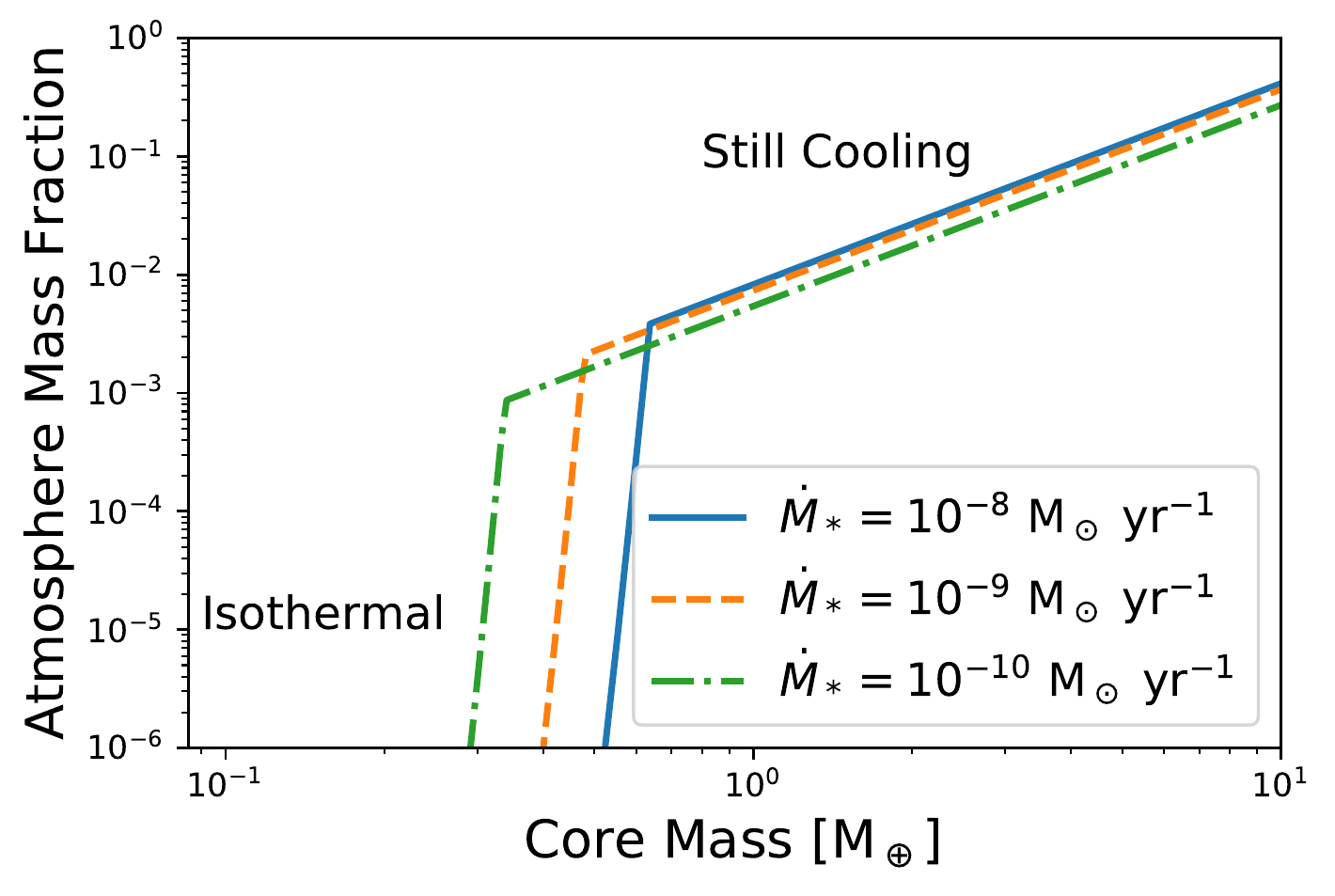}
    \caption{Accreted atmospheric mass fraction ($M_{\rm atm}/M_{\rm core}$) after 3 Myr as a function of planetary mass. The planet is located at 1~AU around a sun-like star. The three lines represent three different accretion rates in the background disc. This figure has been produced considering the analytic core-accretion calculations of \citet{Lee2015} and is similar to their Figure~4.}
    \label{fig:Atmos_acc}
\end{figure}

The envelope accretion models appropriate for terrestrial mass planets that accrete from the nebula are those in which the atmosphere can cool, and as it cools it contracts, allowing more material to become bound to the planet \citep{Ikoma2006,Lee2015,Ginzburg2016,Stokl2016,Jankovic2019}. In these accretion models an adiabatic interior is connected to the nebula via an approximately isothermal radiative zone \citep{Rafikov2006}. The cooling rate (and hence accretion rate) is set by the conditions at the radiative-convective boundary rather than the nebula conditions \citep[e.g.][]{Ikoma2006}. Thus, the resulting atmospheres are quite insensitive to the nebula conditions, but do depend on the optical properties (and hence opacity) of the atmospheric gas. Results of an atmospheric accretion calculation are shown in Figure~\ref{fig:Atmos_acc}. To calculate the accreted mass fractions, we use the analytic core accretion scalings of \citet{Lee2015}. The \citet{Lee2015} analytic models assume ISM-like dust in the upper atmosphere and that the radiative-convective boundary is set by hydrogen dissociation. We consider the accretion of H/He by a planetary core at 1~AU around a sun-like star. The properties of the protoplanetary disc are taken from the model of \citet{Garaud2007} for typical T Tauri star accretion rates. 

Figure~\ref{fig:Atmos_acc} shows that there are two limits to the accretion of an atmosphere. At low core masses, the atmosphere mass is so low that it can cool entirely until it is in radiative equilibrium with the background nebula, and as such is isothermal with the same temperature as the nebula. However, the mass of an isothermal atmosphere increases exponentially with the planetary mass. This is because the density at the Bondi-radius must be of order the disc density and an isothermal atmosphere has an exponential density profile, where the density at the core is roughly:
\begin{equation}
    \rho_{\rm core}=\rho_{\rm nebula}\exp\left(\frac{R_B}{R_{\rm core}}-1\right)\,.
\end{equation}
Since $R_B\propto M_p$, $R_{\rm core}$ is roughly $\propto M_p^{1/4}$ \citep[e.g.][]{Lopez2014} and the atmosphere mass is dominated in the last few scale heights near the core, one can expect $\rho_{\rm core} \propto \exp(M_p^{3/4})$. Eventually, the atmosphere becomes so massive that it can no longer reach radiative balance with the nebula in typical disc-lifetimes. This means there is a net heat flux from the planet's interior into the nebula.  Here the majority of the atmosphere is convective and the envelope is still cooling and has a mass that depends in a power-law fashion to the core mass. Figure~\ref{fig:Atmos_acc} shows that this transition occurs around a core mass of 0.3-0.7~M$_\oplus$ and an atmospheric mass fraction of $\sim0.001-0.005$ for planets located at 1~AU around a solar-mass star. 

The main takeaway of these calculations is that if a protoplanet reached a mass of $\sim 1$~M$_\oplus$ before the gas disc dispersed, then it will accrete an atmospheric mass of approximately 1\% of the core's mass. However, if it remained at lower masses then the atmosphere mass it would accrete from the nebula would be lower, with a 0.4~M$_\oplus$ only accreting an atmosphere mass of only $\sim 10^{-5}-10^{-4}$ of the core's mass in a disc with a typical mass accretion rate of $10^{-9}$~M$_\odot$~yr$^{-1}$. In essence, below about one Earth-mass there is a rapid transition from being able to accrete a negligible H/He atmosphere to accreting a voluminous one. The precise value of this transition does depend on the disc properties. However, once on the ``still cooling'' branch, the accreted atmosphere mass is substantial enough to affect the planet's appearance in a transit survey, yet is fairly insensitive to the disc properties. 

Thus, theoretically, core-accretion models suggest that terrestrial mass planets around $1$~M$_\oplus$ and higher can accrete voluminous hydrogen/helium atmospheres if they reach their final masses before the nebula disperses. 

We add several notes of caution. Firstly, while the protoplanet may be able to accrete these atmospheres, they are still pressure supported by the nebula. In the case of rapid nebula dispersal (as is indicated by observations - e.g. \citealt{Kenyon1995,Ercolano2011}), this rapid loss of pressure support can cause the protoplanets to lose some of the accreted atmosphere \citep{Ikoma2012,OW16,Fossati2017,Ginzburg2018,Kubyshkina2018_young}, which has shown to be significant for $\gtrsim 3$~M$_\oplus$ planets. However, (as indicated in Figure~\ref{fig:Atmos_acc}) the atmospheres of terrestrial mass exoplanets are closer to radiative balance with the nebula and therefore more tightly bound. Since this ``boil-off'' process has not been studied for lower, terrestrial mass planets, it's unclear if this process is important or not. 

Secondly, recent 3D hydrodynamical simulations \citep[e.g.][]{Ormel2015,Fung2015,Cimerman2017} have indicated that the typical core-accretion picture may not be accurate for low-mass protoplanets and that recycling of material between the atmosphere and nebula is an important process. However, simulations by \citet{Kurokawa2018} have also suggested that the impact of recycling may have been overestimated in previous isothermal simulations. Since none of these simulations treats the radiative transfer of the problem properly, it is still unclear how important this recycling might be for proto-atmospheres of terrestrial mass planets. Furthermore, the enhancement of solids in the atmosphere can also modify the accretion of H/He gas \citep[e.g.][]{Hori2011,Venturini2015,Venturini2016,Bodenheimer2018}. In this case the pollution of the atmosphere by solids enhances the atmosphere's mean-molecular weight, thus the mass of accreted atmospheres can be larger. Finally, Kite et al. (2020) has suggested that during the accretion process H/He was mixed and absorbed into a molten core, modifying the accretion process.

\subsection{Indications of nebula gas envelopes around early Venus and Earth}

Growing terrestrial planets can accumulate a H/He atmosphere by accretion of gas from the protoplanetary nebula as discussed above. In particular, inside the habitable zone, this happens if they reach masses that are $\gtrsim 0.5M_{\rm Earth}$ before the disc disperses (Figure~\ref{fig:Atmos_acc}, \citealt{Lammer2014,Stokl2016}). If proto-Venus and proto-Earth accreted such a mass before the disc evaporated away after several million years
\citep[e.g.][]{Owen2011,Bollard2017,Wang2017}, it would have been possible that both planets could have captured a thin H/He atmosphere before they accreted their final mass afterwards. Noble gas arguments in today's Venus and Earth atmosphere point to the fact that noble gases have been trapped from the nebula in solar composition in the interiors and atmospheres \citep{Porcelli2000,Porcelli2001,Yokochi2004,Odert2018}. This evidence indicates that Venus and Earth did indeed accrete some H/He from the solar nebula. 

For Earth, there are data based on measured $^{20}$Ne/$^{22}$Ne isotope ratios from the mantle that indicate that noble gases have also been incorporated in solar composition from the protoplanetary disk in its interior via underlying magma oceans \citep[e.g.][]{Mizuno1980,Porcelli2001,Yokochi2004,Mukhopadhyay2012}. Some researchers, however, argued that these noble gas isotope ratios could have been originated by implantation of the solar wind onto meteors \citep{Raquin2009,Trieloff2000,Moreira2016,Jaupart2017,Peron2017}. This explanation can only explain $^{20}$Ne/$^{22}$Ne ratios of about 12.52-12.75 \citep{Moreira2013,Moreira2016}, but recent studies measured ratios of 13.03$\pm$0.04 from deep mantle plumes and determined a $^{20}$Ne/$^{22}$Ne ratio of 13.23$\pm$0.22 for the primordial mantle \citep{Williams2019}. Since this value is similar to the nebular ratio, these authors suggest this to be solid evidence for a reservoir of nebular gas that preserved in the deep mantle.

Although less data are available for Venus, according to \citet{Cameron1983} and \citet{Pepin1991,Pepin1997}, its high noble gas abundance indicate that the heavier noble gases in the atmosphere are not significantly different from their primordial values. While atmospheric Ne and Ar ratios on Venus are closer to the initial solar ratios, on Earth the atmospheric isotope ratios are different and represent values that are modified by a contribution from carbonaceous chondritic material to the proto-Earth \citep[e.g.][]{Marty2012,Bouvier2016}.

However, on Earth noble gas isotope remnants of the proto-solar nebula are still present in the mantle, indicating a sequestration of nebular gas at an early stage of planetary growth \citep{Porcelli2001,Yokochi2004}. This is confirmed by the discovery of slightly fractionated solar $^{20}$Ne/$^{22}$Ne ratios trapped in fluid inclusions from plume-related rocks from Iceland \citep[e.g.][]{Dixon2000,Porcelli2001} and the Kola Peninsula, in Russia \citep{Yokochi2004}. These findings are also in agreement with the hypothesis of \citet{Pepin1991,Pepin1997,Pepin2000} that some noble gas fraction, which has been found on Earth, such as the so-called U-Xe composition \citep{Pepin2000,Becker2003}, should have been integrated into the planet’s interiors via an early H/He dominated primordial atmosphere getting into contact with a magma ocean. Hence the evidence points towards the fact that Earth and Venus accreted some H/He from the solar nebula. In order to understand how much they could have accreted, we need to understand how H/He atmospheres are retained and lost. 

\section{Models for the loss or retention of H/He atmospheres} \label{sec:loss}
It is apparent from the discussion in the previous section that the accretion of H/He atmospheres by terrestrial mass planets may be a common outcome of planet formation both inside and outside the Solar System. However, since these atmospheres are weakly bound and receive intense amounts of XUV irradiation, particularly when young, they may not retain them indefinitely (as it is clearly the case for the Solar System terrestrials). Since XUV irradiation varies significantly with stellar type \citep[e.g.][]{McDonald2019}, we will consider how atmosphere retention varies with stellar mass. Furthermore, should one desire a planet to become habitable, they will have to eventually lose these primary atmospheres, as the temperature and pressure at the planet's surface with such an atmosphere is not amenable to surface liquid water \citep[e.g.][]{OM16}. 

\subsection{Preliminaries}

We can parse the parameter space and get a sense of where loss and retention of a planet's H/He atmosphere is possible by considering the energy-limited model for mass-loss and considering the timescale for atmosphere loss. Hydrodyanmic escape occurs where high-energy stellar photons are absorbed at large altitudes in a planet's atmosphere, heating it up to high temperatures (several thousand to tens of thousand of Kelvin). These high temperatures can be reached in the upper atmospheres due to the low densities, and lack of strong molecular coolants that dominate deeper in the atmosphere where the bulk of the stellar spectrum is absorbed.  

The energy-limited mass loss rate can be written as \citep[e.g.][]{watson1981,Lammer2003,Baraffe2004}\footnote{For simplicity, we chose to ignore the effect of stellar tides \citep{Erkaev2007} here}:
\begin{equation}
    \dot{M}=\eta \frac{F_{\rm HE}\pi R_p^3}{G M_p}\label{eqn:EL_loss}
\end{equation}
and the mass-loss timescale (i.e. the timescale to lose an atmosphere at the current mass-loss rate) is:
\begin{equation}
    t_{\rm loss}=\frac{M_{\rm atm}}{\dot{M}}\label{eqn:tmdot}\,.
\end{equation}
Here, $\eta$ is the heating efficiency, $F_{\rm HE}$ is the stellar XUV (< 912\,\AA) flux, $R_p$ and $M_p$ are the planetary radius and mass, respectively, and $M_{\rm atm}$ is the atmospheric mass. \citet{Owen2017} showed that for the atmosphere mass of concern here ($M_{\rm atm}\lesssim 0.02 M_{\rm core}$) the mass-loss timescale decreases as the atmosphere is lost (as the total atmosphere mass decreases). Thus, the timescale of concern is whether the mass-loss timescale for the initial atmosphere mass is shorter than the timescale available for mass-loss to occur. 

The loss of hydrogen/helium atmospheres is driven by XUV heating, and the XUV heating is highest when stars are young. Thus, the mass-loss timescale of interest becomes the time span along which the star remains in its high or ``saturated'' phase of XUV output $t_{\rm sat}$. Once the ``saturation'' phase finishes the XUV output of the star drops so fast that any planet's atmosphere that remains at this stage can be retained (\citealt{Owen2017}, though mass loss continues throughout the whole main-sequence life time of the host star \citealt{Kubyshkina2019a,Kubyshkina2019b}). \citet{McDonald2019} construct an empirical relation for how $L_{\rm HE}/L_{\rm bol} \times t_{\rm sat}$ scales with stellar mass finding an approximately $\propto M_*^{-2}$ dependence. The origin of this trend is simply the longer pre-main-sequence lifetimes of lower mass stars, as well as the fact that they spin-down slower. Now, of course, there is scatter from star-to-star even at the same mass \citep{Tu2015}; however, we can consider the trend that at fixed bolometric insolation the effect of mass-loss will be more important around lower-mass stars. 

Therefore combining Equations (\ref{eqn:EL_loss}) \& (\ref{eqn:tmdot}) we find that the atmosphere mass that can be striped scales as:
\begin{equation}
    M_{\rm atm}\propto \eta \frac{F_{\rm bol}}{\rho_p M_*^2}\,, \label{eqn:Matm_remove_Mstar}
\end{equation}
where $\rho_p$ is the density of the planet (solid + atmosphere) and $F_{\rm bol}$ is the bolometric flux. Thus, the dependence of XUV activity on stellar mass means that, all other things being equal, a planet is much more likely to lose its atmosphere around a lower-mass star. 

\subsection{Towards sophisticated mass-loss models}

Simple prescriptions for the atmospheric escape rate (e.g. the energy limited model, Equation~\ref{eqn:EL_loss}) are not adequate for describing escape for planets covering the wide range of observed characteristics \citep[e.g.][]{Kubyshkina2018_grid,Kubyshkina2018_approx}. The most important characteristics of planetary outflows from strongly irradiated exoplanets are total mass-loss rate, temperature of the thermosphere, density and ionization degree as well as typical velocities. These values are needed to analyze how the planetary outflow interacts with the stellar plasma environment and to interpret transit observations. 


The properties of planetary outflows are described by aeronomy models, the essential feature of which is plasma photo-chemistry based on hydrogen and helium, with inclusion of trace elements such as Carbon, Oxygen, Nitrogen, and Magnesium. The other essential feature of these calculations is heating of upper atmosphere by ionizing XUV radiation, possibly taking into account the specific stellar spectra and its wavelength-dependent absorption. Complete absorption of ionizing radiation takes place at sufficiently high pressures where the atmosphere of most exoplanets consists of molecular components. Thus, aeronomy model should resolve upper atmospheres from molecular dominated heights to distances where components are mostly ionized and expand at supersonic velocities, with dissociation and ionization layers in between.

Such fully aeronomy models dedicated to exoplanets are rare and almost exclusively focused on hot Jupiters \citep[e.g.][]{Yelle2004,GarciaMunoz2007,Koskinen2007,Koskinen2013}. It may seem unusual to discuss such models in a review about atmospheric escape from terrestrial planets; however, it is important to realise that the majority of our direct observational probes of exoplanet atmospheric escape are mainly confined to close-in gas giants. To study the hydrodynamics of atmospheric escape, a number of simplified models have also been used, which omit chemistry (though typically hydrogen chemistry is included), components other than hydrogen, spectral composition of stellar radiation and/or some other simplifications \citep[also focusing on close-in giant planets, e.g.][]{Tian2005,Erkaev2005,MurrayClay2009,Kubyshkina2018_grid}. Because of their complexity, all these models are one dimensional, implying that three important, essentially multi-dimensional, aspects of the problem have been either neglected or treated qualitatively: the non-uniformity of planet irradiation, tidal force, and planetary magnetic fields.

For an efficient survey of many systems \citep[e.g.,][]{Lammer2009,Owen2013,Fossati2017,Kubyshkina2018_grid}, a simplified 1D approach enabling fast computing models for calculating mass-loss rates is the best solution. \citet{Kubyshkina2018_approx} even derived an analytical approximation for atmospheric mass-loss rates for planets covering a wide range of parameters based on interpolating the results of thousands of 1D hydrodynamic simulations; they also showed that, particularly for terrestrial planets, this ``hydro-based approximation'' can be several orders of magnitude different to the energy-limited approximation. No fully self-consistent evaluation of 3D effects has been attempted so far, for the simple reason that no 3D models have been developed that include all the necessary physics. The few existing examples of 2D or 3D codes applied to exoplanets are still subject to limitations \citep[e.g.][]{Bisikalo2013,Matsakos2015,Debrecht2018}. 

The first fully 3D multifluid model, which includes aeronomy based on hydrogen photo-chemistry, has been recently developed and used for simulating the hot Jupiter HD 209458b and warm Neptune GJ 436b \citet{Shaikhislamov2018}. This model has been built on the basis of the previous 1D and 2D versions of the same code \citep{Shaikhislamov2014,Shaikhislamov2016,Khodachenko2015,Khodachenko2017}. To get a sense of the comparison between the 1D and 3D models, we can compare in details the outflow simulations for the warm Neptune GJ 436b obtained by the 3D model of \citet{Shaikhislamov2018} with the 1D model of \citet{Loyd2017} (which includes more involved chemistry with Oxygen and Carbon). Both models calculate reactions involving H2, H, H+, H2+, H3+, He, He+.
\begin{figure}
    \centering
    \includegraphics[width=0.75\textwidth]{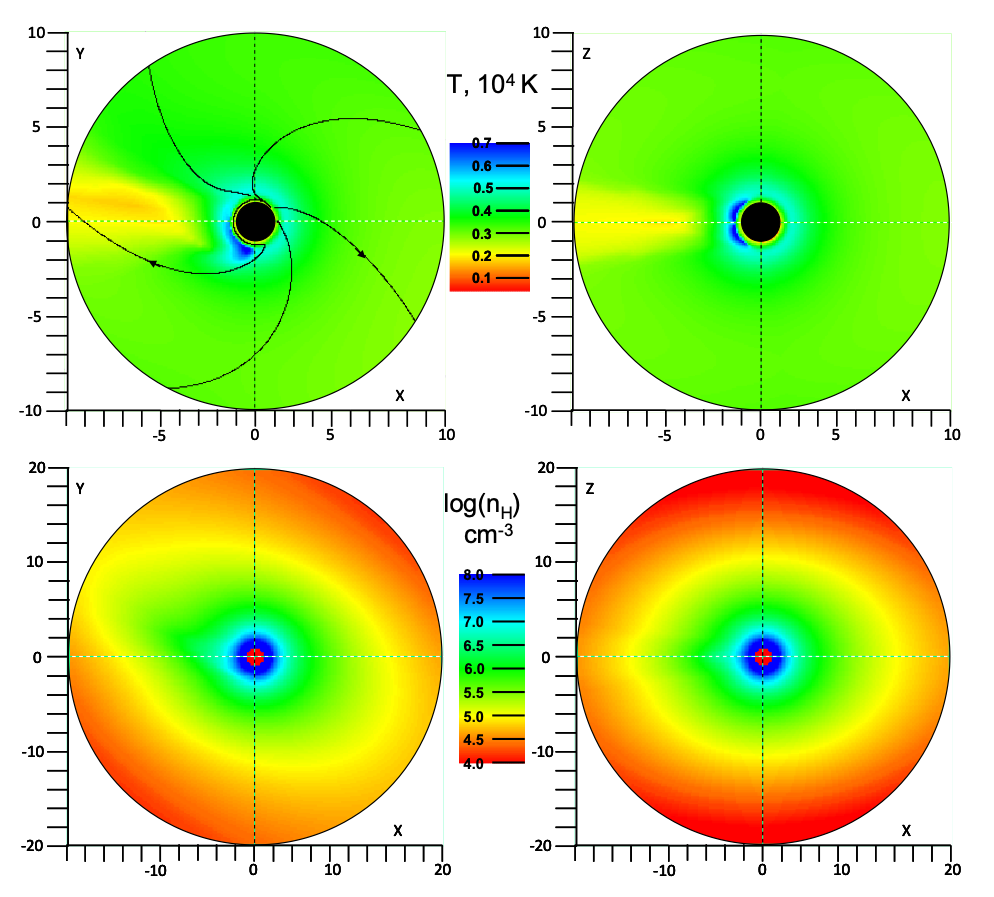}
    \caption{Distribution of temperature (upper row) and density (bottom row) of hydrogen atoms in the equatorial, $X-Y$ (left column), and meridional, $X-Z$ (right column), planes, respectively, calculated for GJ 436b with $F_{\rm XUV}$=0.85 erg s$^{-1}$ cm$^{-2}$ at 1 AU, T$_{\rm base}$=750~K, P$_{\rm base}$=0.03 bar, He/H=0.1. The distance is scaled in units of planetary radii. The star is located on the right at $X=140$ planetary radii. The material velocity streamlines are shown in the top-left panel. The upper panels are limited to $r=10$ planetary radii, whereas the bottom panels cover a twice larger area.}
    \label{fig:3D_GJ436b}
\end{figure}
Figure~\ref{fig:3D_GJ436b} shows the temperature and neutral hydrogen density around the warm Neptune GJ 436b. 
These are cuts along the equatorial and meridional planes and show the 3D structure of the planetary flow. The streamlines show that, in a reference frame where the rotation period of the planet-centred frame equals the orbital period, the material is rotated clockwise by inertia forces as it flows away from the planet. The spiralling of the flow due to the Coriolis force is quite pronounced already at 5 planetary radii. This shows that, with respect to velocity structure, the 1D spherical approximation breaks a few planetary radii away from the planet. For comparison, the sonic point is positioned at about 4 planetary radii. The other 3D effect is a sharp boundary of material distribution along the meridional plane above and below the orbital plane, which is caused by stellar tides. 
Comparisons of these 3D simulations with those of \citet[][their Fig. 4]{Loyd2017} indicate that they are in very good quantitative agreement. For example, in \citet{Loyd2017} the temperature maximum is equal to 4200 K and is positioned at 1.85 planetary radii, the H2 dissociation front is at 1.35 planetary radii, and the outflow velocity reaches a value of 10 km/s at 8 planetary radii. The most important characteristic -- the mass-loss rate -- is equal to about $3\times10^{9}$ g/s in \citet{Loyd2017} and $3.3\times10^{9}$ g/s in the 3D calculation. Thus, for the first time, 3D simulations have indicated that the main quantitative parameters of atmospheric escape from highly irradiated exoplanets can be accurately simulated by 1D models \citep[see also][for the case of the super-Earth $\pi$\,Men\,c]{ildar2020}. This is because the outflow is driven mostly in the upper atmosphere and because the day-side heating by XUV is efficiently redistributed over the whole atmosphere by zonal flows. The maximum of XUV heating rate and the temperature maximum, are located below two planetary radii \citep{GarciaMunoz2007,Shematovich2014,Salz2016}. About 70\% of all XUV radiation in the simulation shown in Fig. 2 is absorbed below 3~$R_p$. Also, the main cooling processes below 3~$R_p$ are H$_3^+$ and Lyman-$\alpha$ emission, while above 3~$R_p$ the cooling by advection and expansion becomes prevailing. Thus, for the generation of planetary atmospheric escape, the actual velocity of the material below 3\,R$_p$ is not particularly relevant. These facts explain why 1D and 3D models agree at distances shorter than several planetary radii.

However, the spatial structure of the outflows at distances beyond several planetary radii, such as flow twisting by Coriolis force and compression by stellar tides, which are certainly important for the interaction with the stellar plasma environment, can only be simulated by 3D models. Therefore, while 1D models can be used to describe the evolution of planetary atmospheres over long timescales (with the caveat of magnetic fields; see below), 3D models are needed for appropriate comparisons with observations.

\subsection{Impact of planetary magnetic fields}

Planetary magnetic fields are yet another complex aspect of the atmospheric escape problem. Planetary magnetic fields interact with both the outflow and the stellar wind. Here we consider the simple, but most relevant question, of the effect of planetary magnetic fields on the outflow of upper atmospheres. Again, since almost all the simulation work focuses on hot jupiters we use these simulations as a guide to our discussion. 

In the first attempts to quantify the importance of magnetic fields, a specific topology of the inner magnetosphere with predefined ``wind-'' (where the outflow occurs along open field lines) and ``dead-zones'' (where the outflow is suppressed within closed field lines) was assumed, and the mass loss of planetary atmospheres was investigated in a semi-analytic way \citet{Trammell2011,Adams2011}. 
Estimates of the size of the ``dead-zone'' and construction of the solution in the ``wind-zone'' made it possible to evaluate the overall mass loss as a function of planetary magnetic field strength, showing that suppressions of an order of magnitude or larger are possible.

Later on, more self-consistent treatments based on 2D MHD codes have been performed \citep{Trammell2014,Owen2014,Khodachenko2015} in which ``dead-'' and ``wind-zones'' have been shown to form in the expanding outflow. Because of the different treatment of thermosphere and heating/cooling processes, the estimated magnetic field at which the mass loss of hot Jupiters is significantly suppressed varied among these papers by about an order of magnitude - from 0.3 G in \citet{Owen2014} up to more than 3 G in \citet{Trammell2014}, and 1 G in \citet{Khodachenko2015}. However, the overall conclusion stands that the atmospheric outflow of exoplanets can be suppressed by a planetary magnetic field, with possible reductions in the mass-loss rates by an order of magnitude \citep[e.g.][]{Khodachenko2012}. However, this conclusion was reached for hot Jupiters such as HD 209458b, but generalising to other exoplanets is not straightforward. 

Based on studies of magnetized analogues of hot jupiters, we can estimate how strong the magnetic fields of other planets should be to significantly suppress the upper atmospheric outflow. Because the plasma is stationary in the dead zone, a balance between gravity and thermal pressure is maintained along closed magnetic field lines. Using available aeronomy models to simulate an upper atmosphere in hydrostatic equilibrium, the temperature and ionization degree are determined by radiative heating/cooling and plasma photo-chemistry. Recently, such approach has been employed to calculate the parameters of upper atmospheres for other types of planets \citet{Weber2018,OA19}. From this analysis, magnetic fields of 0.5-10 G may be able to reduce mass loss from highly irradiated terrestrial mass planets by up to an order of magnitude. The reduction in mass-loss rates due to magnetic fields implies that it would be harder for these planets to lose their primordial atmosphere. However, we emphasise that little direct simulation work has been done in this area. 

\subsection{Models and implications of H/He loss from Earth and Venus}

\citet{Odert2018} applied a hydrodynamical upper atmosphere escape model to proto-Venus atmosphere evolution scenarios. This work found that EUV-driven hydrodynamic escape of a captured H/He atmosphere, corresponding to the XUV activity of a weakly-to-moderately rotating young Sun \citep{Tu2015}, can reproduce the present day observed atmospheric Venus Ne and Ar isotopic ratios. In such scenarios, proto-Venus grew to masses of $\sim$60\% of its final mass inside the gas disk.

The finding of \citet{Odert2018} is also supported by a more detailed and complex study by \citet{Lammer2019}, which included, besides a hydrodynamic upper atmosphere model, outgassing of elements from magma oceans below primordial atmospheres and impact delivery of elements with chondritic elemental ratios. \citet{Lammer2019} found that the present day atmospheric $^{36}$/$^{38}$Ar and $^{20}$/$^{22}$Ne ratios can be reproduced if proto-Venus and Earth were released from the disk with masses of about 0.68-0.81 and 0.53-0.58~M$_{\rm Earth}$, respectively. Thus, Earth's and Venus' early H/He atmospheres were then lost through hydrodynamic escape and possibly impacts of Moon- to Mars-mass planetary embryos \citep{Lammer2019}. In this scenario the escaping H atoms dragged and fractionated the heavier noble gases. According to this study a more massive solar-like H/He atmosphere surrounded early Venus compared to the less massive proto-Earth, which lost its nebula-based primordial atmosphere within a few Myr so that carbonaceous chondrites modified the evolution of the noble gas ratios after the disk dispersal to today's measured atmospheric values, which is reflected in the $^{20}$Ne/$^{22}$Ne ratio.

Furthermore, an interesting finding of \citet{Lammer2019} is that, if proto-Earth would have reached a mass of $\gtrsim 0.7$~M$_{\oplus}$ during the disk life time, a slow to moderate rotating young G-star could not have removed the captured primordial envelope. If a proto-Earth within the habitable zone reached one Earth-mass during the disk life time, it could not have lost its H/He-dominated atmosphere even if the host star originated as a very active fast rotating G-star, simply because it accreted too much H/He to begin with (\S~\ref{sec:acc}). Therefore, a proto-Earth in the habitable zone with a mass in the range 0.7--1.0 M$_\oplus$ habitability depends on the early activity and rotation history of the star. 


\subsection{Models of H/He loss from terrestrial exoplanets}

Atmospheric loss is important for our discussion of the origin of terrestrial planets with secondary atmospheres, and those which could be potentially habitable. As discussed above, it is thought that if the Earth an Venus reached their final masses in the protoplanetary nebula before it dispersed and therefore accreted H/He atmosphere masses ($\sim$ 0.01 M$_\oplus$), the Sun's early XUV output was not sufficiently strong or last long enough to completely remove it \citep{Lammer2019}. Therefore, the evidence points to the fact that Venus and Earth were below their final masses and accreted significantly (exponentially) less H/He, allowing removal. This highlights the importance in understanding whether terrestrial mass planets could accrete and retain voluminous H/He atmospheres in terms of a planets potential habitability \citep[e.g.][]{Lopez2018,Neil2020}. 

\begin{figure}
    \centering
    \includegraphics[width=0.45\textwidth]{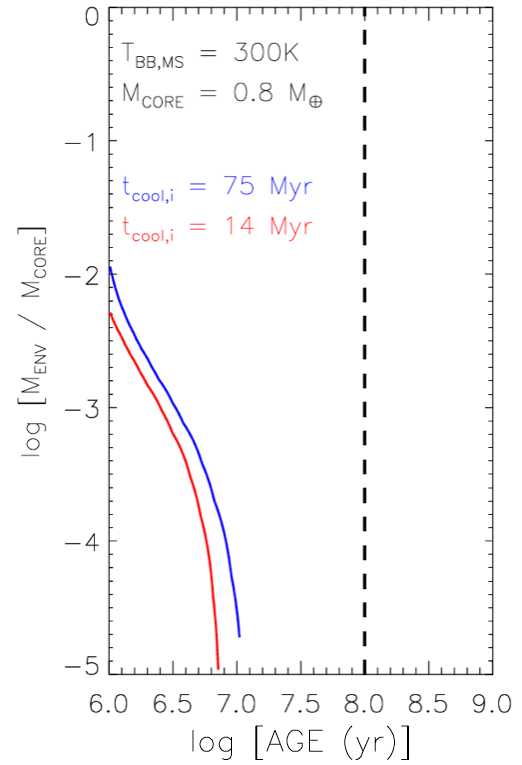}
    \caption{Evolution of 0.8~M$_\oplus$ planets around a 0.4 M$_\odot$ star. The planets are initially released from the protoplanetary nebula with an atmospheric mass of $\sim 1\%$ of the planet's mass and are on orbits such that they would have a black-body temperature of 300~K when the star is on the main-sequence. The read and blue lines show different expected values of the initial atmospheric cooling times (atmospheric entropies; see discussion in \citealt{OM16}), and the vertical dashed line shows the age of the nearby 0.4~M$_\odot$ star AD Leo. Figure taken from \citet{OM16}.}
    \label{fig:OwenMohanty}
\end{figure}

Our questions is of particular importance as one moves to lower mass stars. As demonstrated in Equation~(\ref{eqn:Matm_remove_Mstar}), around lower-mass stars complete removal is possible. In Figure~\ref{fig:OwenMohanty}, we show the evolution of a numerical model of a 0.8~M$_\oplus$ planet with an initial H/He atmosphere of $\sim 1\%$ by mass around a 0.4~M$_\odot$ star (with parameters taking from the nearby star AD Leo). 

\begin{figure}
    \centering
    \includegraphics[width=0.75\textwidth,trim={0 1.35cm 0 0 0}]{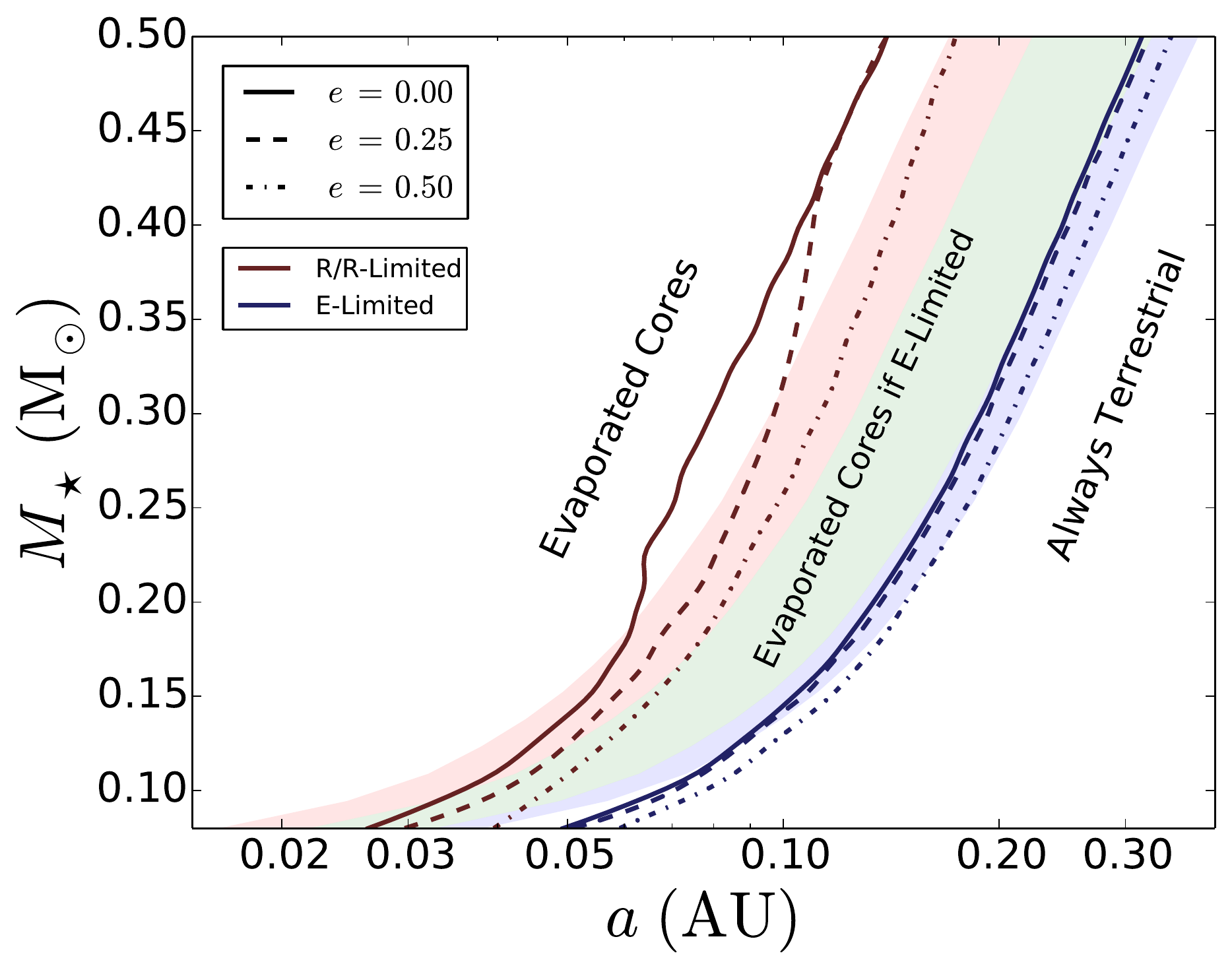}
    \includegraphics[width=0.75\textwidth]{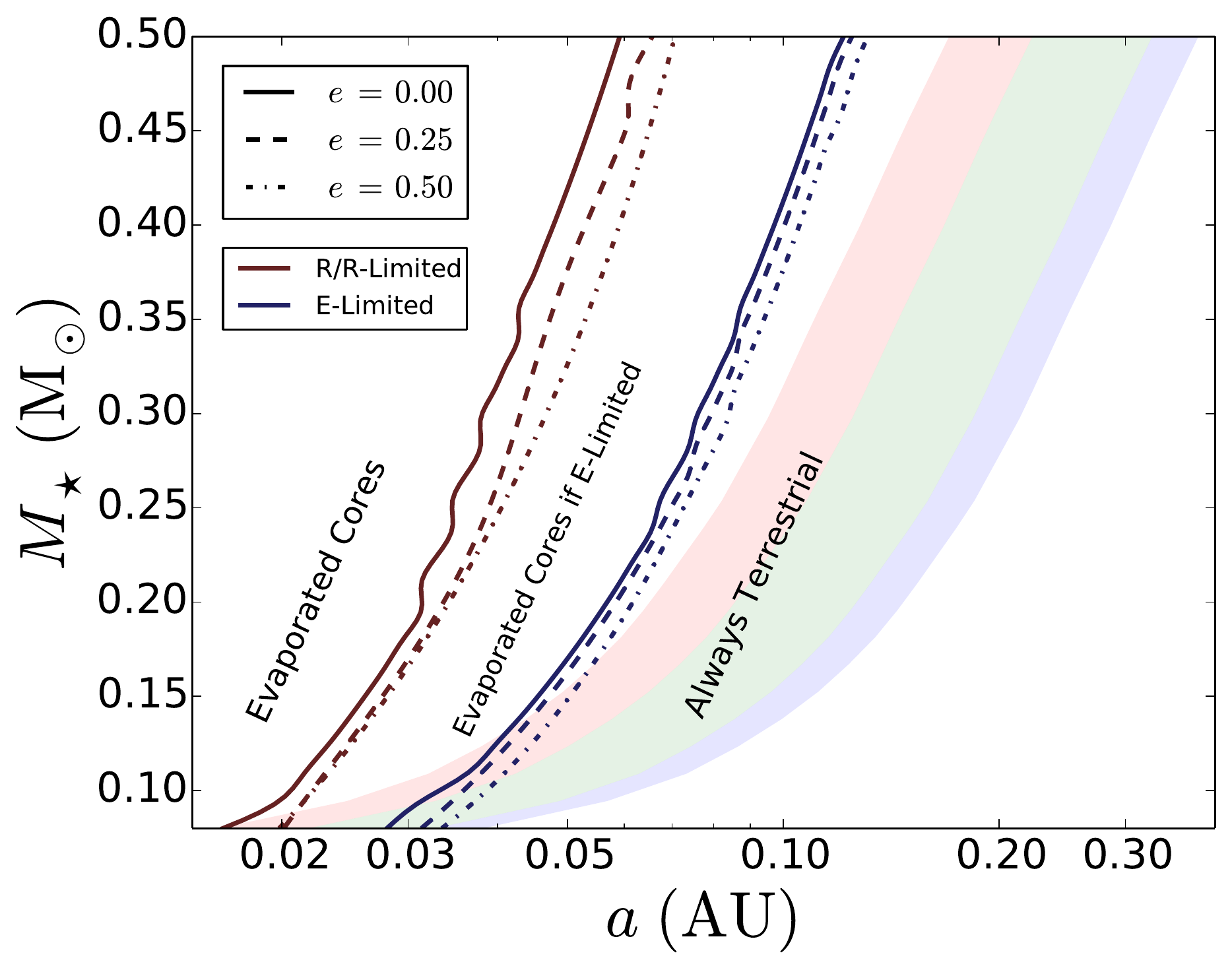}
    \caption{Zones in which 1~M$_\oplus$ planets could not possess a hydrogen/helium atmosphere. Those listed as ``Always Terrestrial'' would need to have been born without such an atmosphere (perhaps after the gas disc dispersed). ``Evaporated cores if energy-limited'' refers to those that could lose an atmosphere if mass-loss was energy-limited and, ``evaporated cores'' are those that could have lost their atmospheres through less efficient mass-loss. The colour bands represent the ``standard'' HZ (green-\citealt{Kopparapu2013}) and an optimistic HZ (red to blue). The different line-styles are for different planetary eccentricities. The top panel is for efficient mass-loss $\eta=0.3$ and a long saturation time of 1~Gyr; whereas the bottom panel is for less efficient mass-loss and a shorter saturation time (100~Myr). Figure taken from \citet{Luger2015}.}
    \label{fig:Luger}
\end{figure}

Therefore, it is entirely possible for habitable zone planets to reach their final terrestrial masses before the nebula disperses though accreting a voluminous H/He atmosphere, which is then lost through escape, but only for stars less massive than the Sun. \citet{Luger2015} explored the parameter space of temperate terrestrial planets around M-dwarf stars using a simple model (i.e., energy-limited approximation). They showed that if mass-loss was highly efficient with $\eta=0.3$,\footnote{Note this is larger than the typical value now found in radiation-hydrodynamic simulations, \citep{Owen2012,Shematovich2014,Salz2016,Erkaev2016,Ionov2018}.} 1~M$_\oplus$ planets in the habitable zones of M-dwarfs would lose their initial H/He atmospheres (top-panel of Figure~\ref{fig:Luger}), whereas a less efficient mass-loss $\eta = 0.15$ around an M-dwarf that had a short saturation time of $100~$Myr would only be able to completely remove a H/He atmosphere in the HZ around the lowest mass ($\sim 0.1$~M$_\odot$) M-dwarfs (bottom-panel of Figure~\ref{fig:Luger}). 

\begin{figure}
    \centering
    \includegraphics[width=0.75\textwidth]{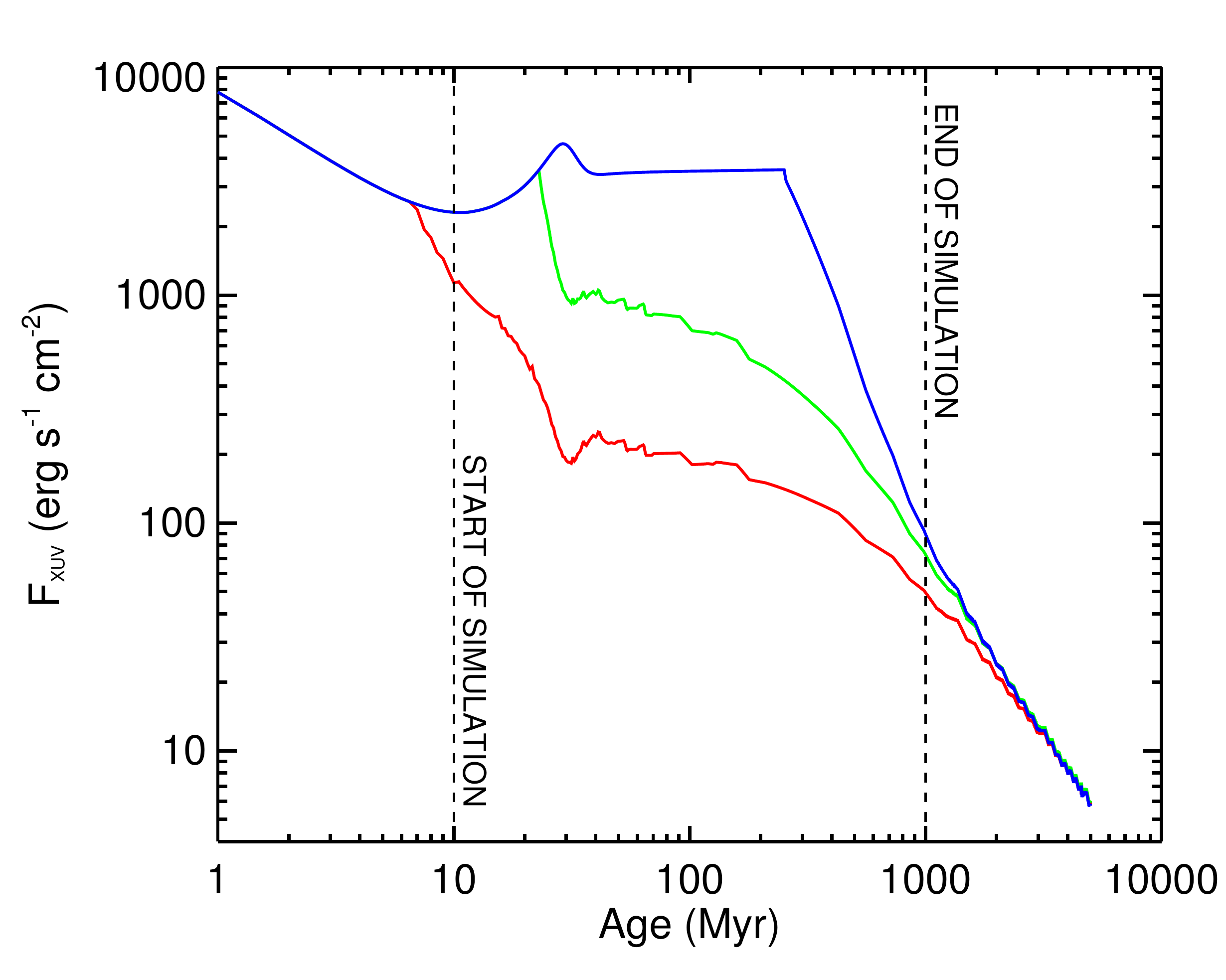}
    \includegraphics[width=0.75\textwidth]{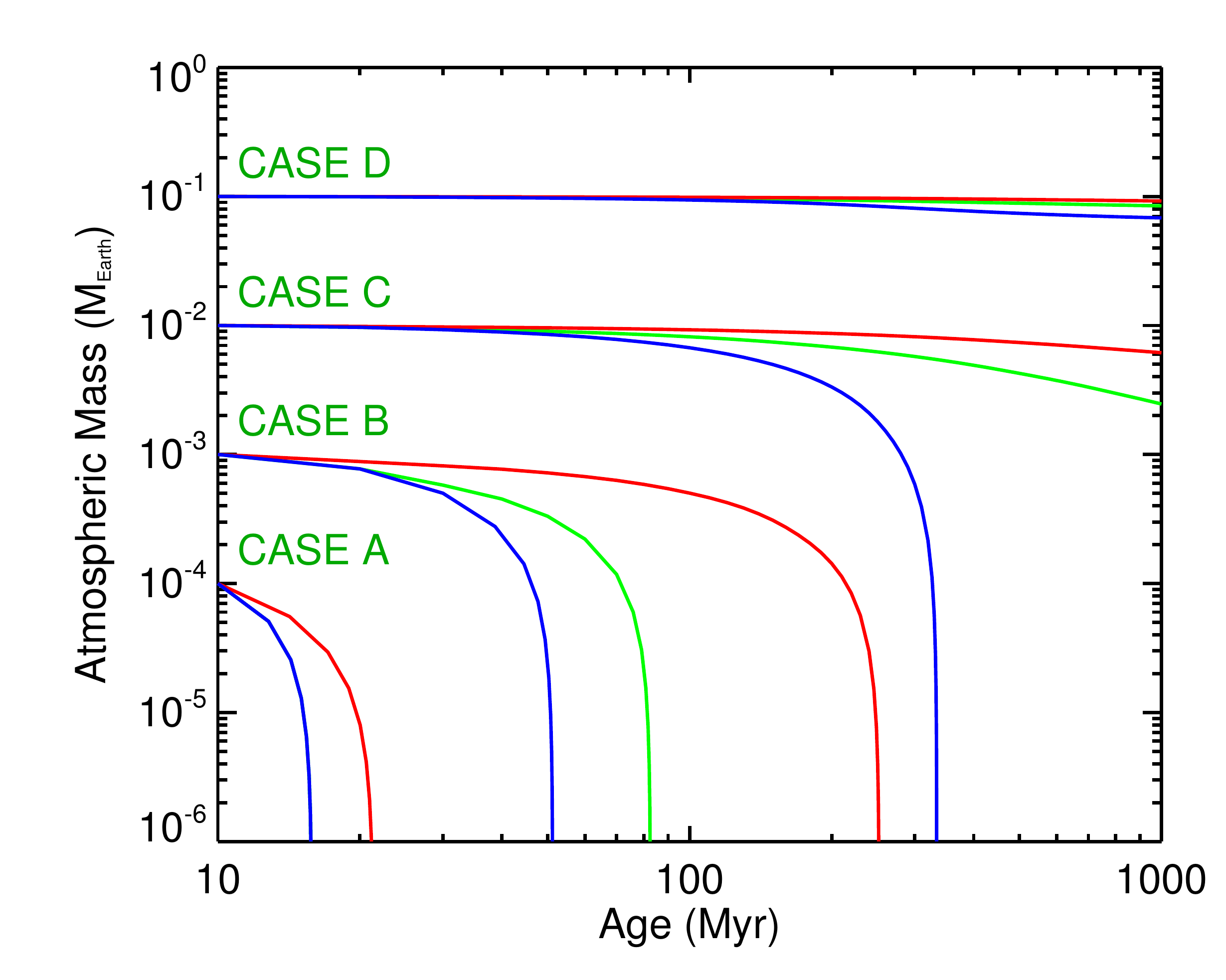}
    \caption{Evolution of the high-energy output of a solar-mass star received by a planet at 1~AU (top panel) for a slow rotator model (red), average rotator model (green) and fast rotator (blue). These XUV fluxes are then used to evolve the atmospheres of 1 M$_\oplus$ planets. Each case shows the evolution of a 1~M$_\oplus$ core orbiting at 1~AU for different initial atmospheric masses (bottom panel).  For a typical $1\%$ atmosphere that would be accreted by a 1~M$_\oplus$ core (Case C) there are diverging evolutionary paths. Figure from \citet{Johnstone2015}.   }
    \label{fig:Johnstone}
\end{figure}

The \citet{Luger2015} results show that not only mass loss efficiency is important, but also the saturation time for the high energy stellar flux is relevant. The evolution of the stellar high-energy flux is coupled to the spin-evolution of the star and therefore the initial stellar rotation rate, where the different initial rotation rates can lead to an approximately order magnitude spread in $L_{\rm sat} \times t_{\rm sat}$ \citep{Tu2015}. \citet{Johnstone2015} investigated this effect, showing that for our typical case of a 1~M$_\oplus$ planet with an initially $1\%$ atmosphere by mass, the spin evolution of the star could either result in the planet retaining the majority of it's atmosphere (Case C - red line in Figure \ref{fig:Johnstone}) or completely losing it (Case C - blue line in Figure~\ref{fig:Johnstone}). \citet{Kubyshkina2019a,Kubyshkina2019b} developed a method for inferring the past rotation evolution of Gyrs-old late-type stars on the basis of the currently observed properties of their close-in sub-Neptune-like planets. This method, further returns a measure of the planetary initial atmospheric mass fraction.

\section{Observational constraints from the exoplanet population}

\begin{figure}
    \centering
    \includegraphics[width=0.75\textwidth]{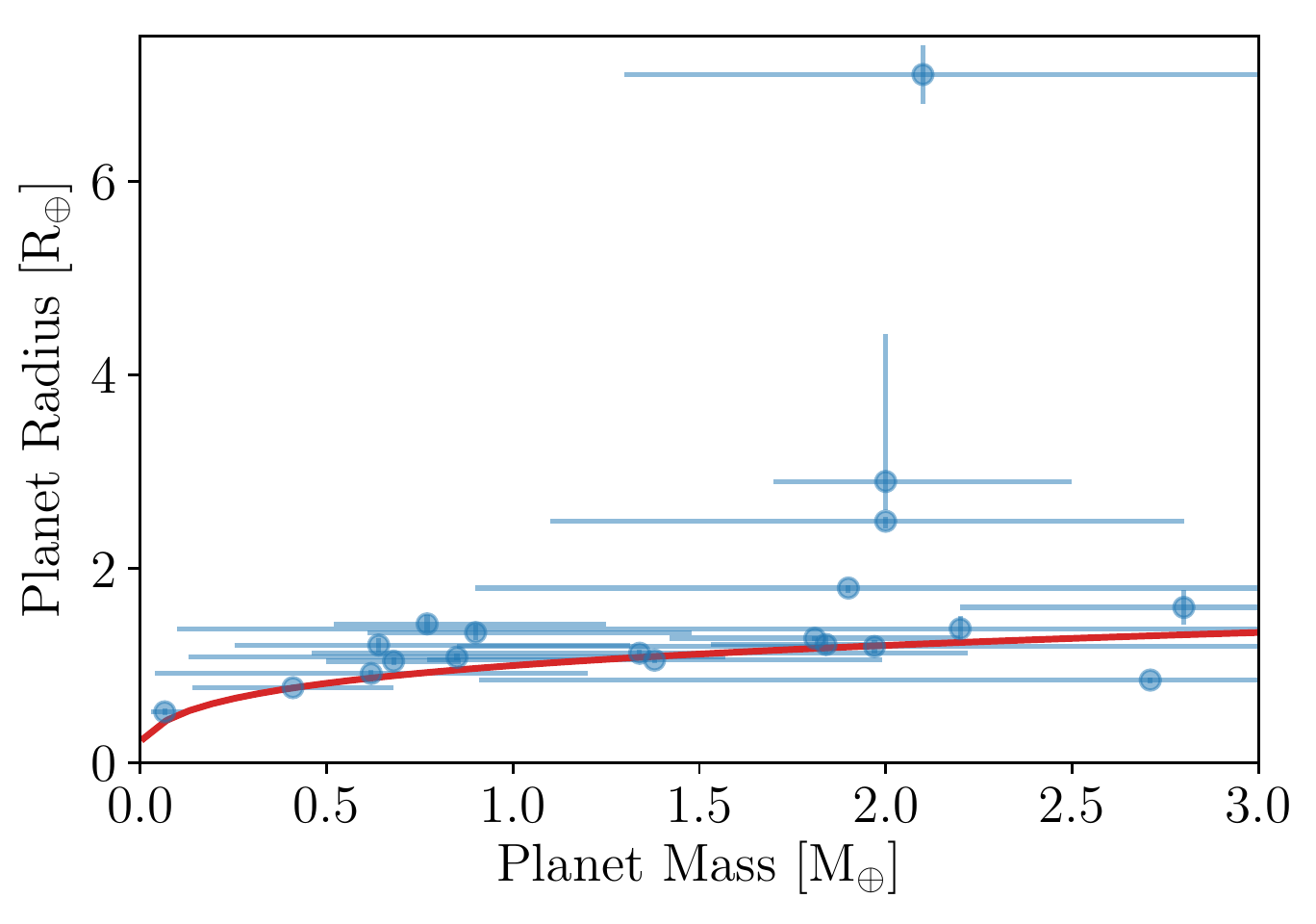}
    \caption{Masses and radii of terrestrial mass exoplanets (those planets with $M_p<3~$M$_\oplus$ and $M_p+\sigma<$~5~M$_\oplus$ are shown). The mass-radius relation for an Earth-like composition is shown, taken from the models of \citet{Fortney2007}. This plot uses data from the NASA exoplanet archive downloaded on 28/10/2019. The lowest mass planets are in the Trappist-1 system.}
    \label{fig:mass_radius}
\end{figure}

Due to observational biases, direct evidence for the accretion and loss of hydrogen/helium atmospheres around terrestrial mass exoplanets is missing. Terrestrial mass exoplanets that can have their masses and radii reliably measured (e.g. {\it Kepler}-78b, \citealt{Howard2013,Pepe2013}) are typically so close to their host stars that models predicted they never retained their atmospheres \citep[e.g.,][]{Owen2018,Kubyshkina2018_grid}.  However, there are a number of $\sim 2$~M$_\oplus$ planets that have radii $\gtrsim 2.5~$R$_\oplus$ indicating they posses and have retained a voluminous H/He atmosphere. The masses and radii of detected terrestrial mass exoplanets are shown in Figure~\ref{fig:mass_radius}. 

This figure shows that, while the vast majority of the lowest mass exoplanets are consistent with a terrestrial, Earth-like composition within the measurement uncertainties (note the vast majority are so close to their stars that models predict they would have lost any initial H/He atmosphere), three planets stand out with significantly larger radii than would be expected if they did not possess a voluminous H/He atmosphere. These planets are {\it Kepler}-11f \citep{Lissauer2011}, {\it Kepler}-51b \citep{Masuda2014} and {\it Kepler}-177b \citep{Xie2014}. These so called ``super-puff'' \citep{JontofHutter2016} planets (because of their extremely low densities -- $\rho_p\lesssim 0.1$g~cm$^{-3}$) are the possible example of terrestrial mass exoplanets that accreted large H/He atmospheres which they then retained \citep[see][for the case of the Kepler-11 system]{Kubyshkina2019b}. All these planets are at long periods ($\gtrsim 40$~days) and their masses have been measured through Transit Timing Variations (TTV) \citep[e.g.][]{Lithwick2012,Wu2013,Hadden2014,JontofHutter2016}. We note that there is a ``lucky'' combination of being detected in a region of low detection efficiency and being in the ``correct'' type of planetary system for the TTV method to be able to measure their masses. This fact implies that there could be many more of these super-puffs than indicated in the data. While $\sim$2~M$_\oplus$ is the lowest mass of an exoplanet known to host a voluminous H/He atmosphere, it is possible that they may exist at even lower masses. 

Probing whether terrestrial mass planets did, in general, form before the nebula dispersed and accreted large H/He atmospheres would require analysing the population of small planets at periods $\gtrsim 100$~days, where models would predict these planets might retain their primary atmospheres. At this distance from their host stars, mass-measurements become very difficult with current instrumentation. Alternatively, if terrestrial planets did not reach their final masses before the nebula dispersed and continued accretion afterwards, they should appear as $1~R_\oplus$ planets in transit surveys. However, {\it Kepler}'s detection efficiency was $<10\%$ for transiting 1 R$_\oplus$ planets with periods longer than 100~days \citep{Fulton2017}. The {\it TESS} mission will not probe these long-periods for the majority of the sky \citep{Barclay2018}. Thus, we must wait for the {\it PLATO} mission for better constraints on the abundance of long-period terrestrial sized exoplanets.   

We can further point to models of loss and retention that attempt to fit the close-in exoplanet population \citep{Owen2017,Jin2018,Wu2019,Gupta2019}. All these works can explain the majority of the population of terrestrial mass exoplanets without a significant H/He atmosphere today through a mechanism in which they accreted one that was then lost. While the constraining data is weak for planets with radii $\sim 1$~R$_\oplus$ \citep{Owen2017}, it is certainly intriguing that current mass-loss models require almost {\it all} observed low-mass exoplanets to have accreted a significant H/He atmosphere. 

Finally, there is tentative evidence of a few small terrestrial exoplanets that could not have lost a significant H/He atmosphere through hydrodynamic atmospheric escape. These planets {\it Kepler}-100c \& 142c could have formed like the Solar-system terrestrials wherein they were lower than their final mass and only accreted a small amount of H/He before undergoing atmospheric escape and collisional growth \citep{Owen2020}.

\section{Discussion and Summary}

The Solar-System terrestrial planets and the exoplanet population appear to point to two differing hypotheses about the accretion and retention of primordial H/He atmospheres. The Solar-System planets show evidence for the accretion of a small amount H/He from the nebula which was then lost. There is also a strong line of evidence indicating that the Solar-System terrestrials were not at their final masses before the nebula dispersed (particularly for the Earth), limiting the amount of H/He they could have accreted, which only becomes significant (in terms of altering the radius and density of the planet) above $\gtrsim 1$~M$_\oplus$. The accretion of a small amount of H/He then makes atmospheric loss efficient during the early active phase of the Sun. 

However, the exoplanet population appears to point to an alternative evolutionary history, in which most planets reached close to their final masses before the nebula dispersed, accreting significant amounts of H/He into large atmospheres that inflated the planet's radius of several Earth radii. Those planets, which are irradiated strongly enough, are able to lose these atmospheres, resulting in terrestrial like planets after a few 100 Myr. Whereas those that are not strongly irradiated enough retain these voluminous H/He atmospheres for Gyrs. Furthermore, the higher {\it relative} XUV output from lower-mass stars implies that mass-loss is more important for planets around stars with masses lower than solar at the same bolometric irradiation level. This stellar mass dependence implies that, if accretion of voluminous H/He atmospheres is dominant for terrestrial mass planets, then HZ rocky exoplanets would only be possible around M-dwarf stars. 

This dichotomy between the information provided by exoplanets and the Solar System planets may not be real. Due to observational biases in the exoplanet data, the parameter spaces between the exoplanet population and the Solar System terrestrial planets do not overlap, with the exoplanet population being well characterised only inside the orbit of Mercury around Sun-like stars. The relative probability of those planets born with voluminous H/He atmospheres and those that formed like the Solar-System terrestrials is likely to be answered by the {\it PLATO} mission, which will finally probe the exoplanet population in the terrestrial range at longer periods compared to what done by {\it CoRoT}, {\it Kepler} and {\it TESS}. 

We also note that considerable theoretical work is still required to understand atmospheric accretion and loss around terrestrial mass exoplanets. On the accretion side, 3D hydrodynamic simulations appear to be giving qualitatively different results to standard core-accretion calculations. Additionally, we must move beyond simple energy-limited models of atmospheric escape, e.g. as different choices of the heating efficiency can lead to quite divergent answers as to the retention and loss of H/He atmospheres around terrestrial mass planets (e.g. Figure~\ref{fig:Luger}, \citealt{Luger2015}). 

Finally, as detection and characterisation of exoplanet atmospheres becomes possible in the coming decade, the exoplanet field must follow the lead of the Solar System terrestrials and calculate the impact H/He loss would have on the abundance of the heavy elements left behind. While measuring isotopic ratios is certainly out of immediate reach, measuring the abundances of heavy elements in the atmospheres for a statistical sample of terrestrial planes may be possible with {\it ARIEL} \citep{Tinetti2018}. Additionally, comparisons between exoplanets in the same systems - especially those systems which contain both planets that have a H/He atmosphere today, and those that may or may not have lost one (e.g. {\it Kepler}-36, \citealt{Carter2012,OM16}) will be crucial for understanding the origin and evolution of terrestrial mass planets.

\begin{acknowledgements}
The authors are grateful to participants of the ISSI workshop on exoplanet atmospheres in November 2018 where this review was initiated. We thank the three anonymous reviewers for comments which improved the manuscript. JEO is supported by a Royal Society University Research Fellowship and a 2019 ERC Starting Grant (PEVAP). IFS is supported by Russian Science Foundation, project 18-12-00080. HL and MLK acknowledge also support by the FWF NFN sub-projects S11606-N16 and S11607-N16. This research has made use of the NASA Exoplanet Archive, which is operated by the California Institute of Technology, under contract with the National Aeronautics and Space Administration under the Exoplanet Exploration Program.
\end{acknowledgements}

%
%

\bibliographystyle{spbasic}      
\bibliography{bibfile}   

%
%

\end{document}